\documentclass[traditabstract]{aa}
\usepackage{txfonts,textcomp}
\usepackage{verbatim}
\usepackage[dvips]{graphicx}
\usepackage{natbib}
\bibpunct{(}{)}{;}{a}{}{,} 

\title{Molecular gas in NUclei of GAlaxies (NUGA)}
\subtitle{XV. Molecular gas kinematics in the inner 3\,kpc of NGC\,6951\thanks{Based on observations carried out with the IRAM Plateau de Bure Interferometer and 30m telescope. IRAM is supported by INSU/CNRS (France), MPG (Germany) and IGN (Spain).}}

\author{T.P.R. van der Laan \inst{1}
\and E. Schinnerer \inst{1}
\and F. Boone \inst{2,3}
\and S. Garc\'{\i}a-Burillo \inst{4}
\and F. Combes \inst{5}
\and S. Haan \inst{6}
\and S. Leon \inst{7}
\and L. Hunt \inst{8}
\and A.J. Baker \inst{9}
}

\institute{Max-Planck-Institut f\"ur Astronomie, K\"onigstuhl 17, 69117 Heidelberg, Germany; \textit{vdlaan@mpia.de}
\and Universit\'{e} de Toulouse; UPS-OMP; IRAP;  Toulouse, France 
\and CNRS; IRAP; 9 Av. colonel Roche, BP 44346, F-31028 Toulouse cedex 4, France
\and Observatorio Astronomico Nacional (OAN)-Observatorio de Madrid, Alfonso XII, 3, 28014-Madrid, Spain
\and LERMA, Observatoire de Paris, 61 Av. de l'Observatoire, 75014-Paris, France
\and Spitzer Science Center, California Institute of Technology, Pasadena, CA 91125, USA
\and Joint ALMA Observatory/ESO, Av. El Golf 40, Piso 18, Las Condes, Santiago, Chili
\and INAF-Osservatorio Astrofisico di Arcetri, Largo E. Fermi 5, 50125 Firenze, Italy
\and Department of Physics and Astronomy, Rutgers, the State University of New Jersey, 136 Frelinghuysen Road, Piscataway, NJ 08854, USA
}

\date{Received 23 March 2010 / Accepted }

\abstract{Within the NUclei of GAlaxies project we have obtained IRAM PdBI and 30m $^{12}$CO(1-0) and $^{12}$CO(2-1) observations of the spiral galaxy NGC\,6951. Previous work shows that there is indirect evidence of gas inflow from 3\,kpc down to small radii: a large-scale stellar bar, a prominent starburst ring (r$\approx$580\,pc) and a LINER/Seyfert 2 nucleus. In this paper we study the gas kinematics as traced by the CO line emission in detail. We quantify the influence of the large-scale stellar bar by constructing an analytical model of the evolution of gas particles in a barred potential. From this model gravitational torques and mass accumulation rates are computed. We compare our model-based gravitational torque results with previous observationally-based ones. The model also shows that the large-scale stellar bar is indeed the dominant force for driving the gas inward, to the starburst ring. Inside the ring itself a nuclear stellar oval might play an important role. Detailed analysis of the CO gas kinematics there shows that emission arises from two co-spatial, but kinematically distinct components at several locations. The main emission component can always be related to the overall bar-driven gas kinematics. The second component exhibits velocities that are larger than expected for gas on stable orbits, has a molecular gas mass of 1.8$\times$10$^{6}$M$_{\sun}$, is very likely connected to the nuclear stellar oval, and is consistent with inflowing motion towards the very center. This may form the last link in the chain of gas inflow towards the active galactic nucleus in NGC\,6951.}
\keywords{Galaxies: individual: NGC\,6951 - Galaxies: nuclei - Galaxies: ISM - Galaxies: Seyfert}

\begin{document}
\maketitle

\section{Introduction}
Gas accretion onto supermassive black holes (SMBHs) is believed to be the cause of nuclear activity in galaxies. SMBHs are accepted as a common component in most galaxies with a significant massive bulge \citep[e.g.][and references therein]{2005SSRv..116..523F}. However, only about half (43\%) of the local galaxies host active galactic nuclei (AGN) in the Seyfert, LINER or transition-object categories \citep{1997ApJ...487..568H}. This discrepancy between the presence of SMBHs and nuclear activity must therefore be sought in the possibility and efficiency of gas transportation to the central regions. Gas transport toward the centers of galaxies can only happen when the gas is able to lose its angular momentum. Two categories of dynamical mechanisms can cause inflow. Gravitational mechanisms such as galaxy-galaxy interactions on large scales or non-axisymmetries (i.e. spiral density waves, large-scale stellar bars or nuclear ovals) within the galaxy potential exert gravitational torques. Viscosity torques and shocks caused by turbulence in the interstellar medium (ISM) are a hydrodynamical mechanism for losing angular momentum. When present, gravitational torques are more efficient \citep[e.g.][]{Combes2004b}.

Observations of the inner regions of galaxies are necessary to understand how the interplay of mechanisms results in gas transportation all the way down to the nucleus. The dynamic timescales in the central region of galaxies are short. Therefore, the gas distribution in these regions needs to be mapped with high angular resolution. The NUclei of GAlaxies (NUGA) project \citep{NUGAstart} has been obtaining high-resolution ($0.5\arcsec - 1\arcsec$) detailed mapping of the molecular gas kinematics in 12 nearby ($D = 4-40$ Mpc) low luminosity active galactic nuclei (LLAGN) with the IRAM  PdBI (Plateau de Bure Interferometer) and 30m telescope. This sample spans the whole sequence of nuclear activity types. In the central kiloparsec most of the gas is in the molecular phase, making CO lines optimal tracers of the gas dynamics. The spatial resolution ($<100$ pc) of this survey allows one to observe the gaseous distribution over an impressive spatial range. This has already led to the identification of a wide range of morphologies in the nuclear regions of these galaxies, including lopsided disks (NGC\,4826: \citet{2003A&A...407..485G}, NGC\,3718: \citet{2005A&A...442..479K}, NGC\,5953: \citet{Casasola2010}), bars and spirals (NGC\,4569: \citet{Boone2007}, NGC\,2782: \citet{Hunt2008}, NGC\,6574: \citet{Lindt-Krieg2008}, NGC\,4579 \citet{Garcia2009}) and rings (NGC\,7217: \citet{Combes2004}, NGC\,3147: \citet{Casasola2008}, NGC\,1961: \citet{Combes2009}).

Large-scale stellar bars are believed to be efficient in driving gas towards the inner Lindblad resonance (ILR) \citep[i.e.][]{Regan2004, 1999Apj...525..691S, 2005ApJ...632..217S}. There they induce the formation of spiral or ring structures \citep{Prendergast1983, Athanassoula1992a, Athanassoula1992b,Englmaier2000, Maciejewski2004II, Combes2004b}. \citet{Martini1999} and \citet{Regan1999} have proposed that spiral shocks induced by the large-scale stellar bar can generate further gas inflow across the ILR. Numerical simulations have shown that spiral structure can extend across the ILR into the nuclear region, if the sound speed is high enough \citep{Englmaier2000} or if the velocity dispersion in the ISM is large enough \citep{Maciejewski2004II}. 

Alternatively, other mechanisms on smaller scales could take over gas transportation towards the nucleus, i.e. the bars-within-bars scenario \citep{Shlosman1989}. Viscosity torques can only become the dominant mechanism for inflow in the innermost regions ($<$200\,pc: \citet{Combes2004b}).

In this paper we study the nearby spiral galaxy NGC\,6951. Its distance of 24.1\,Mpc allows for high spatial resolution ($1\arcsec = 117$pc). There are several arguments that gas inflow is currently occurring in this galaxy. A large-scale stellar bar has been detected in the near-infrared \citep{Buta2003,Block2004} with a bar radius of 26$\arcsec$ ($\approx$3.0\,kpc) and a position angle of 84$\degr$ \citep{Mulchaey1997}. This galaxy also shows a pronounced starburst (SB) ring at 5$\arcsec$ ($=$580\,pc) radius in H$\alpha$ \citep{Perez2000, Rozas2002} and radio \citep{Saikia2002} emission. \citet{Haan2009} have shown that the inner region of NGC\,6951 is HI depleted, implying the ISM must be dominated by its molecular phase. But $^{12}$CO(1-0) and $^{12}$CO(2-1) emission associated with the SB ring has been found \citep{Kohno1999, Garcia2005}, as well as HCN(1-0) emission \citep{Krips2007}. NGC\,6951 has also been classified as `a high excitation LINER and a possibly high nitrogen abundant Seyfert 2' galaxy \citep{Perez2000}. In the central 1$\arcsec$ of NGC\,6951 \citet{Krips2007}  and \citet{Garcia2005} have detected HCN(1-0) and $^{12}$CO(2-1) emission, respectively, indicating that further inflow of gas beyond the SB ring must be or has been occuring recently.

We present $^{12}$CO(1-0) and $^{12}$CO(2-1) observations made with the IRAM PdBI and 30m telescope. Previous papers by \citet{Garcia2005} and \citet{Haan2009} presented the PdBI-only data. Here the $^{12}$CO(1-0) and $^{12}$CO(2-1) data cubes have been combined with the 30m observations. The addition of the 30m observations provides the full CO emission present, sampled on all scales in the inner 3\,kpc of NGC\,6951.

This paper has three aims. The first is to investigate how the addition of the 30m data changes the maps and the gravitational torque results derived by \citet{Garcia2005} and \citet{Haan2009} and to perform a more detailed investigation of the kinematics of the molecular CO gas than has been done before. The second is to quantify the influence of the large-scale stellar bar within the inner 3\,kpc using a parametric kinematic model. The third is to study the impact of the nuclear stellar oval on the gas flow inside the circumnuclear gas ring.

In \S2 we present the observations from the IRAM PdBI and 30m telescope and their reduction. \S3 contains a discussion of the changes due to the addition of the 30m data and a presentation of the spatial and kinematic properties of the CO emitting gas as seen in the PdBI+30m data cubes. In \S4 we detail the large-scale stellar bar model, with the results that follow from the model being presented in \S5. Finally, in \S6 we compare our result with previous gravitational torque studies of NGC\,6951 and discuss observational evidence for inflow to the nucleus. We conclude in \S7.

\section{CO observations}
\subsection{IRAM PdBI observations}
The IRAM PdBI observations in ABCD configuration were carried out between June 2001 and March 2003 using the 6-antenna array in dual-frequency mode. Only the D-configuration observations were executed with 5 antennas. The correlator was centered at 114.726\,GHz and 229.448\,GHz at 3mm and 1mm, respectively, corresponding to a heliocentric velocity of 1425\,km\,s$^{-1}$. The bandwidth covered changed between the CD and AB configurations, although all four configurations covered at least the central $\pm$200\,km s$^{-1}$. The flux calibration used CRL618, 0932+392, 3C273 and/or 3C345. Bandpass correction was derived from observations of a strong quasar at the beginning of the track. 1928+738 and 2010+723 served as the phase calibrators, allowing for correction of atmospheric effects. Phase corrections derived for the 3mm receiver were applied to the 1mm band resulting in a better phase correction at 1mm. The data were reduced using standard routines in the GILDAS software package\footnote{http://www.iram.fr/IRAMFR/GILDAS}.

For both data sets CLEANed data cubes were produced with natural and robust\footnote{Robust weighting here means the weighting function $W(u,v)$ in a \textit{uv} cell is set to a constant if the natural weight is larger than a given threshold, and $W=1$ otherwise.} weighting. Cleaning was done down to the 2$\sigma$ noise level within a fixed polygonal area that was defined based on the zeroth moment map for all channels with line emission. The r.m.s. noise for these data cubes is listed in Table \ref{table:beamsizesPdBI}. 
\subsection{IRAM 30m observation}
30\,m observations of the central 132$\arcsec $ by 66$\arcsec $ were obtained on December 24 and 25, 1997. The 3mm and 1mm receivers were tuned to 114.730\,GHz and 229.460\,GHz. A bandwidth of about 1200\,km/s [600\,km/s] was covered by 512 channels with a width of 2.6\,km/s [1.6\,km/s] at 3mm [1mm]. The spacing between individual grid points was 11$\arcsec$, i.e. half the size of the 3mm beam. The integration time per scan was usually 4\,min, and both polarizations were simultaneously observed for each frequency. Typical system temperatures during the observations were 350\,K and 630\,K for the 3mm and 1mm receivers. The data reduction was done using the GILDAS/CLASS software package. Each scan was inspected and those few with extremely high system temperatures or other instrumental effects were discarded. The baseline was corrected in the individual spectra by fitting a first order polynomial through channels outside the expected line emission. After this correction all spectra for an individual position were averaged together using a noise weight.
\subsection{Short spacing correction}
The 30\,m observations were used to compute the short spacing correction (SSC) and recover the large-scale low-level flux. The 30m observations were reprojected to the field center and frequency of the PdBI observations. The bandwidth coverage of the 30m observations was resampled to match the velocity axis of the interferometric observations. A combined data cube was produced using the task `UV-short' in GILDAS. The single dish weight scaling factor was 6.55$\times$10$^{-3}$ for $^{12}$CO(1-0) and 7.74$\times$10$^{-4}$ for $^{12}$CO(2-1).

Two sets of final CLEANed data cubes were produced using natural and robust weighting. The resolution of the natural [robust] weighted $^{12}$CO(1-0) and $^{12}$CO(2-1) data is given in Table \ref{table:beamsizesSSC}. The data cubes have 512 x 512 pixels, with a pixel scale of 0.25$\arcsec$/pixel [0.10$\arcsec$/pixel] and velocity bins of 10 [5] km\,s$^{-1}$ for the $^{12}$CO(1-0) [$^{12}$CO(2-1)] data. The r.m.s. noise per channel in the SSC $^{12}$CO(1-0) observations is 2.5 mJy/beam and 2.2 mJy/beam with natural and robust weighting, respectively. For the SSC $^{12}$CO(2-1) observations these values are both 7.8 mJy/beam. CLEANing was done down to the 2$\sigma$ noise level with the assistance of individual polygons defined for each channel with line emission present.

The noise values of the SSC data cubes are, with the exception of the $^{12}$CO(2-1) natural weighted data cube, below the noise levels of the PdBI only data. The beam sizes increase slightly ($\sim$ 8\%) due to the added short spacings.
\begin{table}
\begin{minipage}{\columnwidth}
\caption{Overview of PdBI data}
\label{table:beamsizesPdBI}
\begin{tabular}{c l c c c c}
\hline\hline
Emission Line & Weighting & Beam Size & PA (\degr) & r.m.s.\\
 & & ($\arcsec$ x $\arcsec$) & & (mJy/beam)\\
\hline
$^{12}$CO(1-0) & natural & 2.56x1.70 & 108 & 2.8\\
 & robust & 1.37x1.08 & 114 & 2.3\\
$^{12}$CO(2-1) & natural & 1.58x1.37 & 87 & 7.8\\
 & robust & 0.64x0.50 & 111 & 8.0\\
\hline
\end{tabular}\\

\textbf{Notes:} The beam sizes and PAs of the CLEAN beams for the PdBI data. The right most column gives the r.m.s. noise of the CLEANed PdBI-only data cubes. The data cubes are specified according to emission line and weighting.
\end{minipage}
\end{table}
\begin{table}
\begin{minipage}{\columnwidth}
\caption{Overview of PdBI+30m data}
\label{table:beamsizesSSC}
\begin{tabular}{c l c c c}
\hline\hline
Emission Line & Weighting & Beam Size & PA (\degr) & r.m.s.\\
 & & ($\arcsec$ x $\arcsec$) & & (mJy/beam)\\
\hline
$^{12}$CO(1-0) & natural & 3.11x2.59 & 94 & 2.5\\
 & robust & 1.57x1.22 & 112 & 2.2\\
$^{12}$CO(2-1) & natural & 1.72x1.56 & 81 & 7.8\\
 & robust & 0.69x0.55 & 113 & 7.8\\
\hline
\end{tabular}\\

\textbf{Notes:} The beam sizes and PAs of the CLEAN beams for SSC (PdBI + 30m) data. The right most column gives the r.m.s. noise of the CLEANed SSC (PdBI + 30m) data cubes. The data cubes are specified according to emission line and weighting. 
\end{minipage}
\end{table}
\subsection{HST/NICMOS observations}\label{sec:HST}
We retrieved from the HST archive the NICMOS F110W and F160W images of NGC\,6951. The reduction was carried out using the "best" calibration files, and the van der Marel algorithm \citep[e.g.][]{1999ApJS..124...95B} to reduce the "pedestal" effect. Sky values were assumed to be zero, which is generally a good assumption for these kinds of NICMOS images \citep{2004ApJ...616..707H}. The images were calibrated, converted to magnitude scale, and subtracted to obtain a [F110W]-[F160W]$\approx J-H$ color map. This color image and the starburst ring it reveals will be discussed in Sect. 3.1.
\section{Properties of the CO-emitting gas}\label{sec:properties}
\subsection{Morphology and H$_{2}$ masses}\label{sec:morph}
\begin{figure*}[htpb]
\centering
\resizebox{\hsize}{!}{\includegraphics{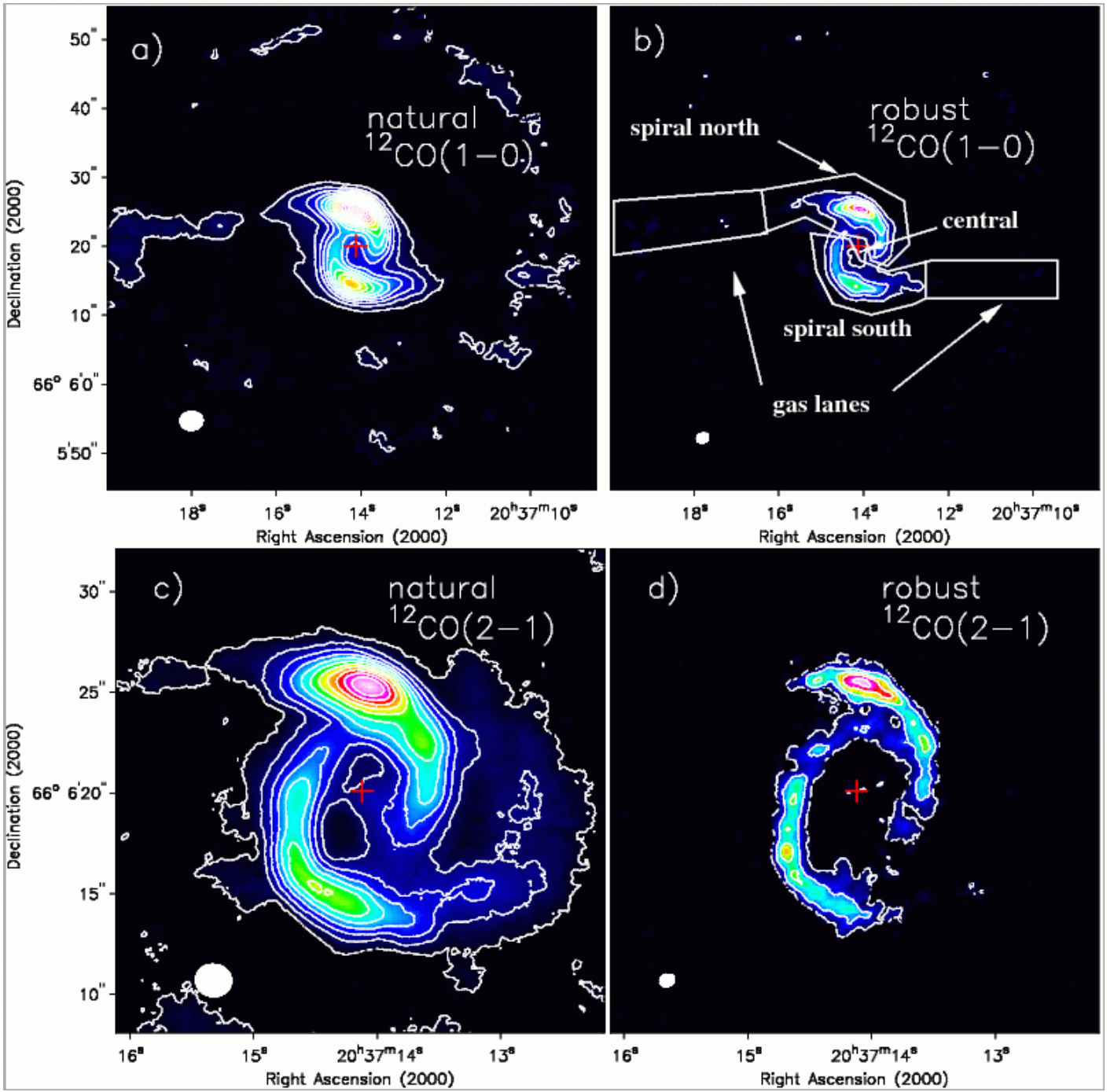}}
\caption{\textit{Top:} Integrated $^{12}$CO(1-0) emission in natural (\textbf{a}) and robust (\textbf{b}) weighting for the SSC data. The zeroth moment maps have been primary beam corrected. The $^{12}$CO(1-0) emission has been integrated from -220 km s$^{-1}$ to 250 km s$^{-1}$. Contours run from 5$\sigma$ in steps of 10$\sigma$. The r.m.s. value is: \textbf{a}) 1$\sigma$ = 0.025 Jy beam$^{-1}$ km s$^{-1}$ and \textbf{b}) 1$\sigma$ = 0.022 Jy beam$^{-1}$ km s$^{-1}$. \textit{Bottom:} Integrated $^{12}$CO(2-1) emission in natural (\textbf{c}) and robust (\textbf{d}) weighting for the SSC data. The $^{12}$CO(2-1) emission has been integrated from -200 km s$^{-1}$ to 200 km s$^{-1}$. Contours run from \textbf{c}) 5$\sigma$ in steps of 20$\sigma$, \textbf{d}) 3$\sigma$ in steps of 10$\sigma$. The r.m.s. value is: 1$\sigma$ = 0.039 Jy beam$^{-1}$ km s$^{-1}$ for both maps. The red cross indicates the position of the dynamical center in all panels (Table \ref{table:theovals}). The beam sizes are shown in the lower left corners and correspond to the values listed in Table \ref{table:beamsizesSSC}.}
\label{CO10intens}
\end{figure*}
\begin{table}[htpb]
\begin{minipage}[t]{\columnwidth}
\caption{Properties of NGC\,6951}
\label{table:theovals}
\renewcommand{\footnoterule}{}
\begin{tabular}{l l l}
\hline\hline
Parameter & Value & Reference\\
\hline
Type & SAB(rs)bc & (1)\\
Nuclear Activity & LINER/Seyfert 2 & (2) \\
\multicolumn{2}{l}{Dynamical Center (Locus Radio Continuum)}&\\
\,\,\,\,RA (J2000) & 20$^h$37$^m$14.123$^s$ & (3,4) \\
\,\,\,\,Dec (J2000) & 66\degr06\arcmin20.09\arcsec & (3,4)\\
Inclination Angle & 46\degr & (5)\\
Position Angle & 138\degr & (5)\\
Adopted Distance & 24.1 Mpc & (6)\\
\hline
\end{tabular}\\

\textbf{References}: (1) \citet{deVaucouleurs}, (2) \citet{Perez2000}, (3) LEDA, \citet{Krips2007}, (4) \citet{Saikia2002}, (5) \citet{Haan2009}, (6) \citet{Tully}
\end{minipage}
\end{table}
\begin{table}[htpb]
\begin{minipage}[t]{\columnwidth}
\caption{CO line fluxes}
\label{table:coflux}
\renewcommand{\footnoterule}{}
\begin{tabular}{l c c c c}
\hline\hline
Component  & \multicolumn{2}{c}{CO(1-0)} & \multicolumn{2}{c}{CO(2-1)} \\
 & natural & robust & natural & robust \\
 & S$_{CO}$ & S$_{CO}$ & S$_{CO}$ & S$_{CO}$\\
\hline
gas lane N & 23.3 & 2.3 & & \\
gas lane S & 10.3 & & \\
spiral north & 142.4 & 123.9 & 280.3 & 151.8\\
spiral south & 109.9 & 94.2 & 202.8 & 123.3\\
central & 10.3 & 6.1 & 22.8 & 0.9\\
\hline
total & 296.2 & 226.5 & 505.9 & 276.0 \\
\hline
\end{tabular}\\

\textbf{Notes}: Integrated line fluxes (S$_{CO}$) for different components of the observed CO morphology. CO fluxes are given in Jy km s$^{-1}$.
\end{minipage}
\end{table}

\begin{table}[htpb]
\begin{minipage}[t]{\columnwidth}
\caption{M$_{H_{2}}$ masses}
\label{table:masses}
\renewcommand{\footnoterule}{}
\begin{tabular}{l c c c c}
\hline\hline
Component  & \multicolumn{2}{c}{CO(1-0)} & \multicolumn{2}{c}{CO(2-1)} \\
 & natural & robust & natural & robust \\
 & M$_{H_{2}}$ & M$_{H_{2}}$ & M$_{H_{2}}$ & M$_{H_{2}}$\\
\hline
gas lane N & 1.2 & 0.11 & & \\
gas lane S & 0.5 & & \\
spiral north & 7.1 & 6.2 & 2.8 & 1.5\\
spiral south & 5.5 & 4.7 & 2.0 & 1.2\\
central & 0.5 & 0.3 & 0.2 & 0.01\\
\hline
total & 15.2 & 11.6 & 5.0 & 2.7 \\
\hline
\end{tabular}\\

\textbf{Notes}: H$_{2}$ masses (M$_{H_{2}}$) for different components of the observed CO morphology. H$_{2}$ masses in 10$^{8}$ M$_{\sun}$. For the CO(2-1) derived masses, we assume I$_{CO(2-1)}$/I$_{CO(1-0)}$ = 0.8.
\end{minipage}
\end{table}
The intensity maps of the CO(1-0) and CO(2-1) observations presented in Fig. \ref{CO10intens} have been constructed from the CLEANed data cubes using the software GIPSY\footnote{Groningen Image Processing SYstem} \citep{1992ASPC...25..131V}. These zeroth moment maps are computed as the pixel-wise sum of emission above a fixed threshold. Here we chose the 3$\sigma$ noise level. Spurious signals are filtered out by imposing the constraint that the emission be above this threshold in at least two consecutive velocity channels. All signals that satisfy these requirements are added to the intensity map. To the CO(1-0) emission zeroth moment maps we have applied a primary beam correction.

The intensity distribution of the observed CO(1-0) and CO(2-1) emitting gas has all the components we would expect from a gas distribution driven by a large-scale stellar bar. The morphology of the CO(1-0) emission map (Fig. \ref{CO10intens}, \textit{top, left}) from larger radii inward is comprised of the following components. At the edge of our field we see a spatially unresolved resonance ring at a radius of $\sim$ 30$\arcsec$ ($=$3.5\,kpc). This is most likely at the location of the ultra harmonic 4:1 resonance inside corotation \citep[e.g.][]{2000ApJ...533..149S}. This ring is connected to straight gas lanes, with an approximately horizontal (east-west) orientation. Their position corresponds well to the orientation of the large-scale stellar bar, which has a position angle of 84$\degr$ \citep{Mulchaey1997}. The gas lanes also coincide with the dust lanes at the leading edges of the stellar bar \citep[][Fig. 4a of their paper]{Perez2000}. In the robust weighted map of the CO(1-0) emission (Fig. \ref{CO10intens}, \textit{top, right}), which is more sensitive to emission on smaller scales, the straight gas lanes are no longer detectable. This indicates that the straight gas lanes consist of more diffuse, low intensity gas.

Going further inward, we find the straight gas lanes connect to a `twin peaks' morphology and a spiral pattern. This was also observed by \citet{Kohno1999}. It is likely that the `twin peaks' are due to the crowding of gas streamlines in a barred potential \citep{Kenney1992}. The orbit crowding leads to a buildup of gas at those locations where orbits change from $x_{1}$ to $x_{2}$, i.e. with orientation along the bar major axis changing to orientation along the bar minor axis for elliptical orbits. The peaks inside the spiral arms have a distance of $\sim$ 6$\arcsec$ ($=$ 700\,pc) from the nucleus. Slightly within that radius ($\sim$ 5$\arcsec$, 580\,pc) a circumnuclear SB ring has been detected in H$\alpha$ \citep{Perez2000, Rozas2002} and radio emission \citep{Saikia2002}.

The natural and robust weighted intensity CO(2-1) maps (Fig. \ref{CO10intens}, \textit{bottom}) show a distribution very similar to the one seen in the CO(1-0) maps. Some of the differences, however, stem from the smaller field-of-view (FOV) of the CO(2-1) data. The outer ring at the edge of the CO(1-0) FOV is not visible, nor are the straight gas lanes. Due to the higher resolution, the two prominent peaks seen in the CO(1-0) maps now break up into multiple maxima. In the natural weighted map the northern peak is now joined by a second one slightly ($\phi\sim$30\textdegree) offset to the west. The southern spiral structure shows an elongated ridge with a far less distinct center. In the natural weighted CO(2-1) map the peaks and ridge are still unresolved structures.
\begin{figure}[tpb]
\resizebox{\hsize}{!}{\includegraphics[angle=-90]{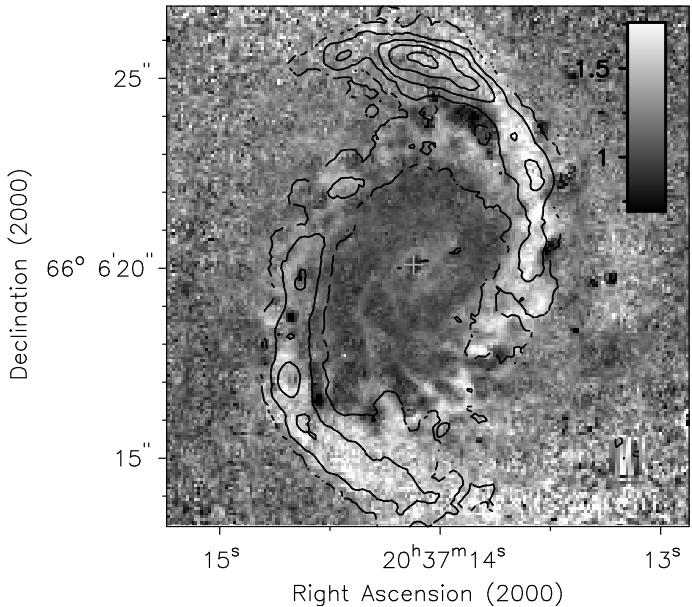}}
\caption{$^{12}$CO(2-1) robust weighted intensity map (contours from 3$\sigma$ in steps of 10$\sigma$ steps, 1$\sigma$ = 0.039 Jy beam$^{-1}$ km s$^{-1}$); overlayed on a J-H color map (greyscale) from \textit{HST}. White corresponds to higher extinction.}
\label{JH-overlay}
\end{figure}
\begin{figure}[htpb]
\resizebox{\hsize}{!}{\includegraphics{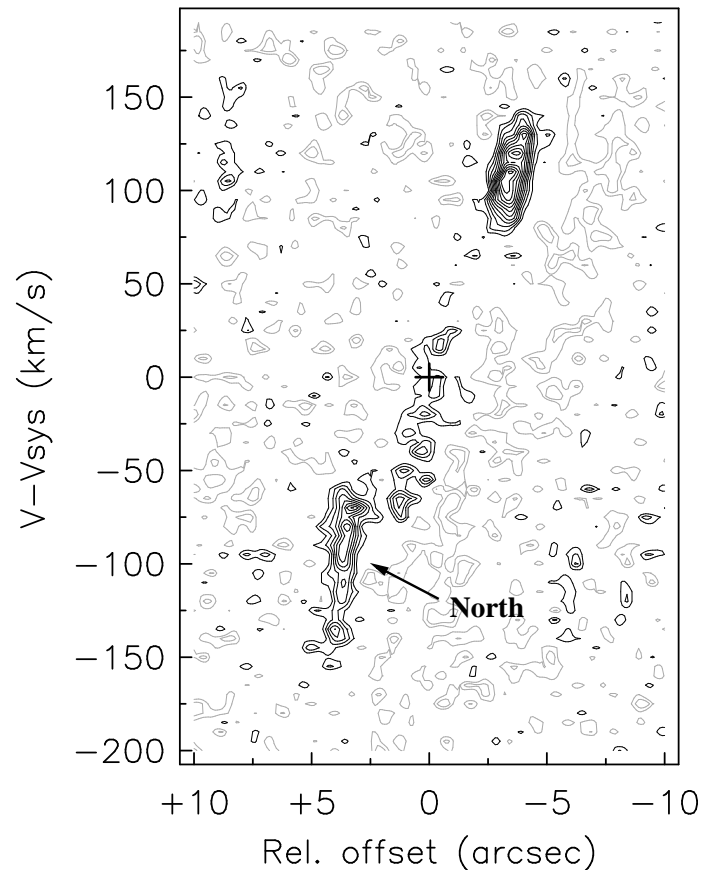}}
\caption{Position-velocity diagram along a PA of 113$\degr$ through the dynamical center. Contours are at 2$\sigma$, in 1$\sigma$ steps, with the -2$\sigma$ and -1$\sigma$ contours given in grey. A gas bridge is visible between the northern part of the ring (lower left here) and the nucleus (denoted with a cross).}
\label{gasbridge}
\end{figure}
\begin{figure}
\resizebox{\hsize}{!}{\includegraphics[width=17cm, angle=-90]{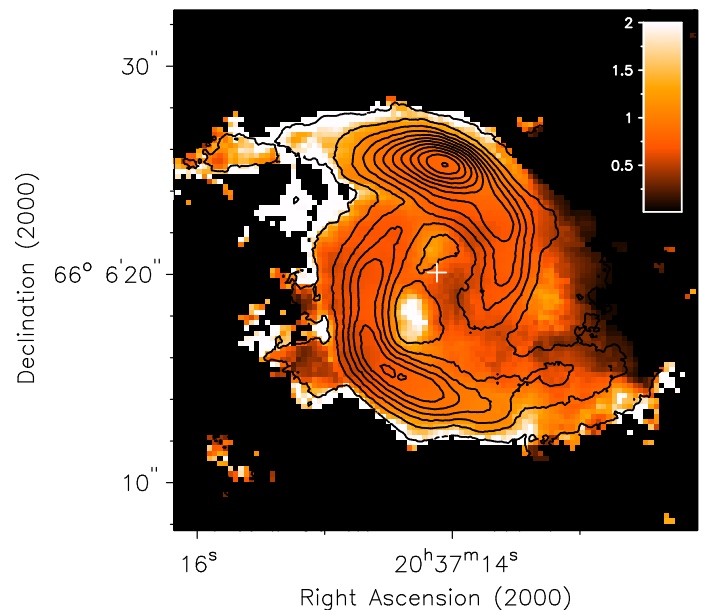}}
\caption{Emission line ratio CO(2-1)/CO(1-0) in temperature units. This ratio was computed using the natural weighted CO(2-1) and the robust weighted CO(1-0) data cubes, smoothed to the resolution of the former. The high ratio south-east of the nucleus inside the ring is not significant.}
\label{linratio}
\end{figure}

We start to resolve two tightly wound gas spiral arms, as well as peaks of dense gas within the ring only in our highest resolution robust weighted CO(2-1) maps. In the higher density tracer HCN(1-0) a similar effect is seen \citep{Krips2007}. The peaks found in that high resolution image, although barely resolved, sit almost 90$\degr$ down the spiral arms. This seems to indicate some density gradient in the gas ring, starting at the northern and southern peaks and going downstream. Fig. \ref{JH-overlay} shows the CO(2-1) integrated intensity map as contours superimposed on the NICMOS $J-H$ color map (Sect. \ref{sec:HST}). The figure shows that the CO spiral arms coincides nicely with the SB ring seen in the red $J-H$ color. The extinction, $A_{\rm V}$, can be derived from this color image assuming a $J-H$ color of $\sim$0.7 for a normal (unreddened) composite stellar population, and ($A_{\rm J} - A_{\rm H}$)/$A_{\rm V}$\,=\,0.092 (as in \citet{Cardelli1989}, with $R_{\rm V}$\,=\,3.1). The $A_{\rm V}$'s so derived for the SB ring are quite red, ranging from $\la$1 to more than 6 in the dustiest, most compact regions.

Finally, at the dynamical center (Table \ref{table:theovals}) we find barely resolved molecular gas emission (M$\approx10^{6}$ M$_{\sun}$) at the 3$\sigma$ level, as already reported by \citet{Garcia2005}. Central molecular gas emission has also been detected in HCN at a much more prominent level \citep{Krips2007}. The gas bridge between the ring and the nucleus, identified by \citet{Garcia2005} as lying at P.A. 134$\degr$, is also observed in our data cubes. Here we find the bridge to be present between PAs of 78$\degr$ and 158$\degr$, but the most intense emission is along a PA 113$\degr$, as shown in the position-velocity diagram in Fig. \ref{gasbridge}.

The morphology discussed here was for the most part also observed in the previously published PdBI-only maps. The exception is the spatially unresolved resonance ring at the edge of our field, which was not seen at all. The straight gas lanes are more prominent in the SSC maps presented here. The consequences for the gravitational torques will be discussed in Section \ref{sec:torquecomp}.

In Tables \ref{table:coflux} and \ref{table:masses} we list the integrated CO line fluxes and the derived H$_{2}$ masses for the different components described above. The regions over which the flux has been measured are indicated in Fig. \ref{CO10intens}b. These values have been derived with a CO-to-H$_{2}$ conversion factor X$_{\rm CO}$ of 2.2 $\times$ 10$^{20}$ cm$^{-2}$ [K km s$^{-1}$]$^{-1}$ \citep{Solomon1991} and have not been corrected for helium abundance. As the CO line ratio (Sect. \ref{sec:ratio}) is fairly constant in the center, the assumption of a single conversion factor seems valid for our purpose. There are claims that the CO-to-H$_{2}$ conversion factor is a factor 3-4 lower in gas-rich centres of spiral galaxies \citep[e.g.][]{2001A&A...365..571W}. A lower conversion factor would lower our mass estimates correspondingly. However, for consistency with other papers in this series, we use the Galactic value. The total integrated $^{12}$CO(1-0) flux within a 40$\arcsec$ field of view is 314 $\pm$ 55 Jy km s$^{-1}$. The $^{12}$CO(1-0) flux we find, is similar to the measurements of 334 $\pm$ 12 Jy km s$^{-1}$ within 65$\arcsec$ by \citet[][obtained with the Nobeyama Millimeter Array and the 45m telescope]{Kohno1999} and 350 $\pm$ 41 Jy km s$^{-1}$ within 45$\arcsec$ by \citet[][with the FCRAO single dish telescope]{1995ApJS...98..219Y}. The corresponding H$_{2}$ mass is $1.6\times10^{9}$ M$_{\sun}$. For the $^{12}$CO(2-1) emission we measure an integrated flux of 452 $\pm$ 65 Jy km s$^{-1}$ within a field of view of 14$\arcsec$, which corresponds to an M$_{H_{2}}$ of $4.6\times10^{8}$ M$_{\sun}$ if we assume I$_{CO(2-1)}$/I$_{CO(1-0)}$ = 0.8 (see Sect. \ref{sec:ratio}). As can be seen from Table \ref{table:masses} the total measured mass in CO is completely contained in the components mentioned. The areas within which flux is measured are the same for natural and robust weighted maps. In both CO(1-0) and CO(2-1) lines we see that the natural weighted maps are more sensitive to low-level diffuse components of the CO gas. The areas over which the flux of the components has been measured, are marked in Fig. \ref{CO10intens}.
\subsection{Line ratio}\label{sec:ratio}
The CO(2-1)/CO(1-0) line ratio has been used to convert the observed CO(2-1) flux into equivalent CO masses in Section \ref{sec:morph}. We have derived the line ratio (Fig. \ref{linratio}) in the following manner. Both maps, CO(1-0) and CO(2-1), have been short-spacing corrected and sample down to the same minimum $uv$ radius. The robust weighted CO(1-0) data cube was smoothed to the resolution of the CO(2-1) natural weighted cube. Then the zeroth moment map of this smoothed CO(1-0) cube was constructed as before. The CO(2-1) zeroth moment map was regridded to the pixel scale of the CO(1-0) map (from 0.10$\arcsec$ to 0.25$\arcsec$). The fluxes of both maps were converted into temperature and the ratio taken.

The ratio is almost constant along the spiral arms, with the map displaying an average ratio of 0.8. The high ratios found at the edges of the ring are insignificant due to low S/N in the CO maps.
\subsection{Kinematics}\label{sec:kin}
\begin{figure*}[htpb]
\centering
\includegraphics[width=17cm]{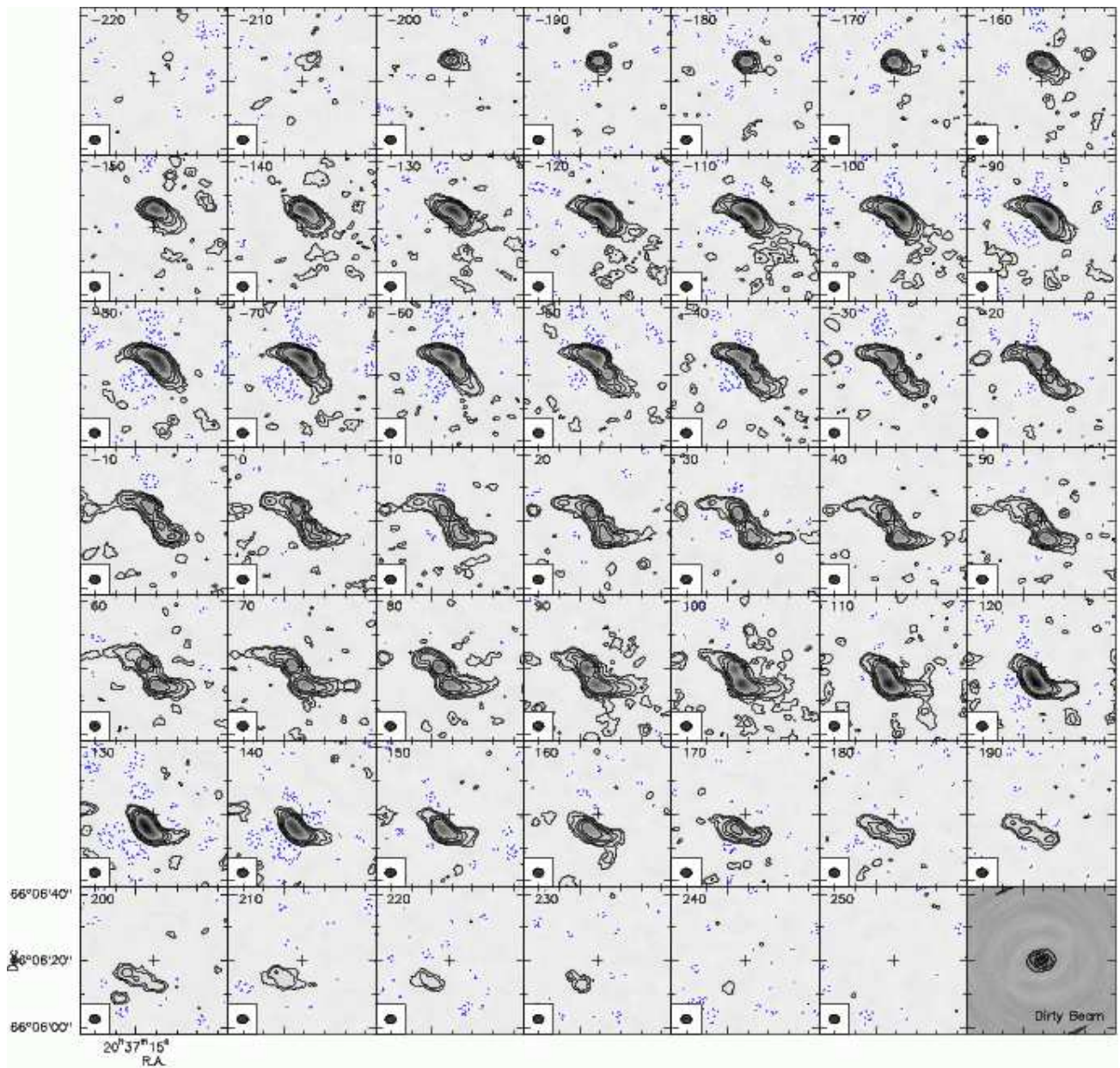}
\caption{Channel maps of the naturally weighted $^{12}$CO(1-0) data cube. The size of each channel map is 44$\arcsec$ by 44$\arcsec$. The contours are at $-$3$\sigma$, $-$2$\sigma$, 3$\sigma$, 5$\sigma$, 10$\sigma$, 15$\sigma$ and 25$\sigma$, with 1$\sigma$ = 2.5 mJy/beam. The velocity relative to systemic velocity of the galaxy (v$_{sys}$ = 1425 km s$^{-1}$) is indicated in the upper left corner. The dynamical center is indicated by a cross in each channel map. The synthesized beam is given in the lower left corner of each channel and the dirty beam is shown in the lower right panel.}
\label{channelmaps_co10}
\end{figure*}

\begin{figure*}[htpb]
\centering
\includegraphics[width=17cm]{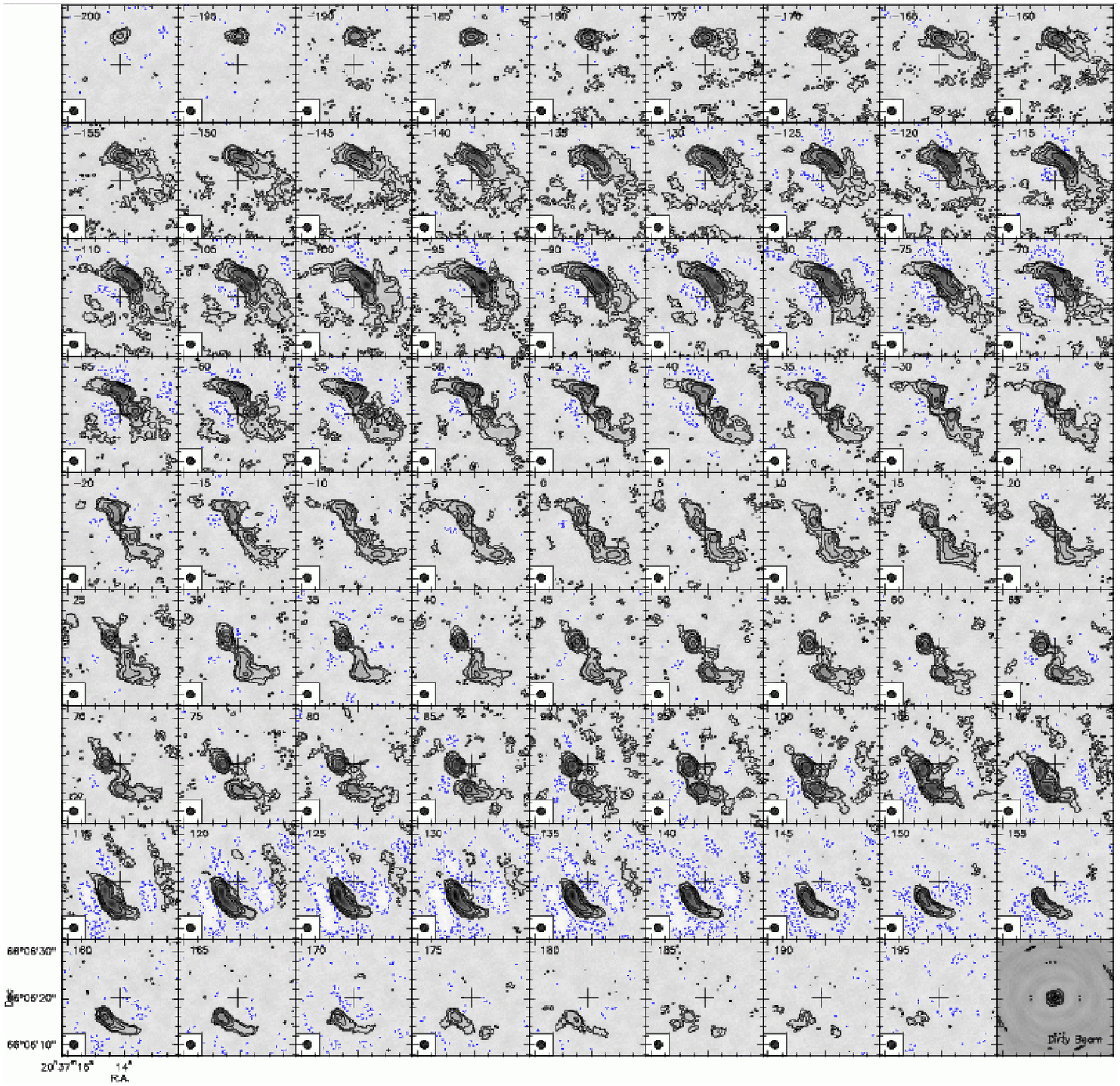}
\caption{Channel maps of the naturally weighted $^{12}$CO(2-1) data cube. The size of each channel map is 25$\arcsec$ by 25$\arcsec$. The contours are at $-$3$\sigma$, $-$2$\sigma$, 3$\sigma$, 5$\sigma$, 10$\sigma$, 15$\sigma$ and 25$\sigma$, with 1$\sigma$ = 7.8 mJy/beam. All other notation is as in Fig. \ref{channelmaps_co10}.}
\label{channelmaps_co21}
\end{figure*}

Evidence of the dominant influence of the large-scale stellar bar is visible in the velocity field. CO(1-0) line emission has been detected in the velocity range of $-$220 km s$^{-1}$ to 250 km s$^{-1}$ relative to the heliocentric velocity of 1425 km s$^{-1}$ in the natural weighted data cube. For the CO(2-1) emission, the velocity range is slightly smaller; from -200 km s$^{-1}$ to 195 km s$^{-1}$. That is in part due to the higher rms noise in this data cube, which affects detection of the signal in the channel maps at the highest relative velocity offsets. All channel maps with significant emission are shown in Figs. \ref{channelmaps_co10} and \ref{channelmaps_co21}. As the velocity increases, the line emission shifts from the north-west to the south-east. This is consistent with the major kinematic axis of this galaxy, measured by \citet{Haan2009} as 138\degr using HI data (Table \ref{table:theovals}). The channels close to systemic velocity ($-$50 to 100 km s$^{-1}$) show two maxima, from both sides of the CO ring, as well as extended arms from the bar-driven straight gas lanes.

\begin{figure*}
\centering
\includegraphics[scale=0.65]{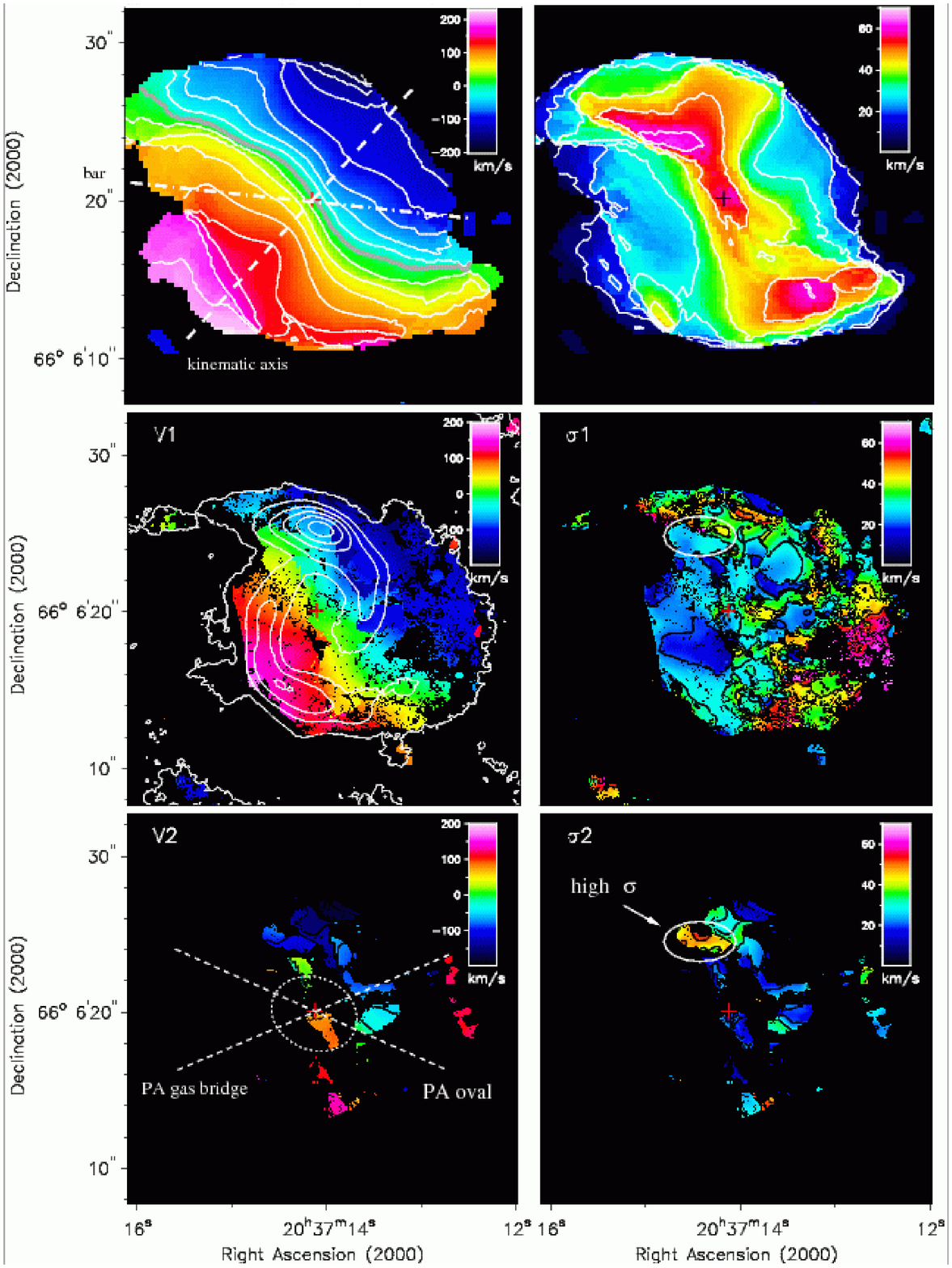}
\caption{\textit{Top left:} First moment map of the CO(1-0) emission. Contours: -200 to 175 km s$^{-1}$ with 50 km s$^{-1}$ steps, 0 km s$^{-1}$ thick gray contour. Kinematic major axis (white dashed line at PA $=$ 138\degr), orientation of the large-scale bar (dashed-dotted line, PA $=$ 84\degr). \textit{Top right:} Second moment map of the $^{12}$CO(1-0) emission. Contours: 0 to 70 km s$^{-1}$ in 10 km s$^{-1}$ steps. The dynamical center is marked by a black cross. \textit{Middle/Bottom panels:} Decomposed velocity field of the CO(2-1) natural weighted line emission. Black spots correspond to blanks due to bad fits at some spatial pixels. \textit{Middle left:} The central value of the Gaussian fitted for each spatial pixel. If two Gaussians where fitted, the value of the Gaussian closest to the large-scale bar model velocity is given. The white contours represent the CO(2-1) intensity map (5$\sigma$ in 40$\sigma$ steps). \textit{Bottom left:} For the spatial pixels where a double Gaussian was fitted, we plot the central value of that second Gaussian here, i.e. the value further away from the bar model velocity. The size and PA of the nuclear stellar oval and PA of the gasbridge, that will be discussed in Sect. \ref{sec:nucflow}, are highlighted with a white ellipse and dashed lines. \textit{Middle/Bottom right:} Velocity dispersion corresponding to respective velocity components. The dispersion is the Gaussian fitted $\sigma$ at each position. Contours from 0 to 70 km s$^{-1}$. The region with significant differences in velocity dispersion (discussed in Sect. \ref{sec:kin}) is highlighted with a white ellipse.}
\label{velomap21}
\end{figure*}

The CO(1-0) velocity map is shown in the top left panel of Fig. \ref{velomap21}. The iso-velocity contours in the center are almost perpendicular to the major kinematic axis. As the radius increases the iso-velocity contours bend, forming the `S'-shape distinctive of velocity fields dominated by large-scale bars.

The dispersion (see the velocity dispersion map in the right panel of Fig. \ref{velomap21}) reaches values of up to 70 km s$^{-1}$. These values are high for a gas disk and have led us to further investigate the kinematics of the CO. The data cubes show that the observed CO line emission arises from two distinct components in velocity space at several positions, most prominently where the dispersion maps show high values. This complex velocity structure means that we are not seeing a truly high {\it local} velocity dispersion in a single component, but rather the projection of two velocity components within the same beam. The assumption of a single component when we determine the velocity dispersion is clearly wrong for some positions. We see these double peaks in both CO(1-0) cubes and the natural weighted CO(2-1) cube.

\begin{figure}[!htpb]
\resizebox{\hsize}{!}{\includegraphics{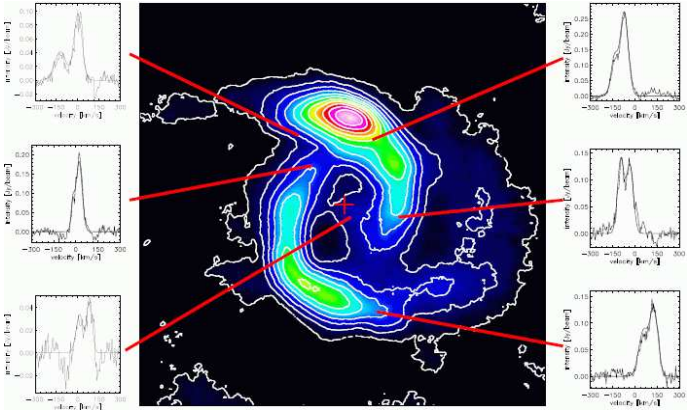}}
\caption{Six example spectra from the robustly weighted CO(2-1) data cube where we fitted a double Gaussian to the velocity axis. The location of each spectrum is indicated with red lines in the intensity map (middle; same area as Fig. \ref{CO10intens} bottom,left panel).}
\label{dub_spec}
\end{figure}

In order to quantify the double emission peaks in detail, we fitted double Gaussians along the spectral axis at each spatial pixel of the CO(2-1) natural weighted data cube (Fig. \ref{dub_spec}). The choice of this cube was made based on the higher spectral resolution of the (2-1) cube with respect to the CO(1-0) cubes. We used the function 'XGAUFIT' from GIPSY for the fitting. At the spatial pixels where we have a double peak, we separate the two components based on the large-scale bar model we derive in Sect. \ref{sec:model}. In Fig. \ref{velomap21} (\textit{middle, left}) we show the central value of the fitted Gaussian for pixels requiring only a single Gaussian fit, and the central value of the Gaussian fit closest to the bar model for the pixels with a double Gaussian fit (component `V1'). This results in a good representation of the velocity field of the disk of the galaxy. In Fig. \ref{velomap21} (\textit{bottom, left}) we present the central values of the second Gaussian (component `V2'). For the most part, the double peaks are present in the northern spiral arm/ring region, with the exception of the complex in the south-south-west close to the nucleus. The velocity difference, $\Delta$v, between the two components is between 40 km s$^{-1}$ and 120 km s$^{-1}$.

Most of the two components might be interpreted in terms of $x_{1}$ and $x_{2}$ orbits. $x_{1}$ orbits are parallel to the large-scale bar (building up the straight gas lanes), and $x_{2}$ orbits are perpendicular to the bar (supporting the ring). At the resolution of the data cube (beam size 1.72'' by 1.56''), the two orbit families might blend together. We find double emission peaks in the CO(1-0) cubes and the natural weighted CO(2-1) cube, where we are unable to resolve the end of the one spiral arm from the other spiral arm. Double velocity components would be the natural consequence of the emission of CO gas on the two orbits families being blended together spatially in our data cubes. In comparison, in the robust weighted CO(2-1) cube we do not find evidence of double peaks in velocity. We find little difference in the width (velocity dispersion) of the Gaussians (Fig. \ref{velomap21} \textit{middle/bottom, right}), except in the northern region. There we see a narrower component, connected to `V1', with a dispersion of about 30 km s$^{-1}$, and a wider component, connected to `V2', with a value of around 40-50 km s$^{-1}$. This region is where we have the northern peak of the `twin peaks' morphology, which persists even in the high resolution CO(2-1) intensity map (in the south we see a ridge). The high dispersion here might be explained by the gas being shocked, which leads to more real, local turbulence/dispersion.

The resolution argument does not explain the double emission peaks found inside the spiral arms, south-south-west of the nucleus. In this region, based on the CO(2-1) map, we expect very little emission, let alone emission arising from two distinct ($\Delta$v $\sim$ 80 km s$^{-1}$) components. The velocity of the first component agrees well with what is expected from the large-scale bar velocity field. The second component's central velocity ($\sim$ 70-90 km s$^{-1}$) seems to connect it with the southeastern part of the ring. There is little difference in dispersion between the two components. The strength of the second Gaussian is nearly equal to the first component. This double emission region is not spatially coincident with the gas bridge mentioned earlier. In Section \ref{sec:nucflow} we will discuss the significance of both the bridge and this second kinematic component in the nuclear region for nuclear fueling.
\section{Kinematic modeling}
\subsection{Motivation}
NGC\,6951 has a prominent large-scale stellar bar with a radius of 3.0\,kpc at a position angle of 84$\degr$ \citep{Mulchaey1997}. We suggested in Sect. \ref{sec:properties} that the gas morphology and kinematics in NGC\,6951 can be well explained in terms of this barred potential. The spiral pattern and the high intensity twin peak morphology are both consistent with gas streamlines under the influence of a large-scale bar.

To further investigate this claim, we model our CO observations with a gravitational potential where the only non-axisymmetric component is a large-scale bar. This cannot be done from observations directly since this would assume exact prior knowledge of the non-axisymmetric components of the gravitational potential. At the same time this modeling will also allow us to derive gravitational torques in this region for an independent comparison with the gravitational torque results of \citet{Garcia2005} and \citet{Haan2009} (Sect. \ref{sec:torquecomp}).

\subsection{Description of the model}\label{sec:model}
We modeled the inner 5\,kpc (diameter) of the observed CO(1-0) gas distribution and its underlying gravitational potential using the DALIA modeling software \citep{2006EAS....18..299B} that was already used to model the NUGA target NGC\,4569 by \citet{Boone2007}.

The tool constructs a kinematic model of gas particles in the presence of a barred potential. The potential is built up from two components, an axisymmetric component representing the disk, and a weak m=2 perturbation: the `bar'. The axisymmetric component has a logarithmic shape and is defined by a characteristic length (r$_{p}$) and velocity (v$_{p}$). 
\begin{equation}
 \Phi_0(r) = \frac{1}{2}v^{2}_{p}ln(1+\frac{r^2}{r^{2}_{p}})
\end{equation}
The bar has the same radial profile as the axisymmetric component, but is tapered by a sine in azimuth and can be oriented in any preferred direction relative to the galaxy's kinematic major axis (Table \ref{barmodel:preamble}: perturbation azimuth). It is further defined by the relative amplitude of the perturbation with respect to the axisymmetric component (i.e. the bar strength, $\epsilon$), and its pattern speed.
\begin{equation}
 \Phi_b(r,\phi) = \epsilon\Phi_{0}cos(2\phi)
\end{equation}
The total kinematic model potential can therefore be described as follows:
\begin{equation}
 \Phi_{total}(r,\phi) = \Phi_0(r) + \Phi_b(r,\phi) = (1+\epsilon {\rm cos}(2\phi))\Phi_0
\end{equation}

The model is populated with a distribution of gas particles, described in terms of the radially varying parameters column density, velocity dispersion and scale height. Two dissipation terms, one acting radially and one acting azimuthally, are set to reproduce the dissipative behavior of the gas particles. Finally the model can be inclined and positioned corresponding to the observations, and a rotation sense is set.

This model has several limitations as discussed by \citet{Boone2007}. First, it is singular at corotation. \citet{Buta2003} and \citet{Block2004} place NGC\,6951's bar radius at 3.0\,kpc and indeed, as can be seen from Fig. \ref{barmodel:figs} (\textit{top, right}), our model does not extend to corotation. Second, closed orbits are computed, which are not self-consistent with the inclusion of two dissipative terms. Therefore, the locations of resonances obtained from this model should be taken as a first-order estimate only.

All parameters in the model can in principle be freely chosen. Our interest here is in obtaining a good set of parameters that reproduces well the observed CO gas distribution. For this reason, we adopted first guesses for the model parameters based on known values of the system and subsequently slightly varied them so that the model better fits our observations. The radial gas mass distribution (Table \ref{barmodel:radii}) was deduced from the CO(1-0) emission maps (Fig. \ref{CO10intens}, \textit{top}) and can be scaled with the total observed H$_{2}$ mass. For the velocity dispersion we take a single value, 30 km s$^{-1}$, which is reasonable for these radii and the assumption of a pressure supported gas disk \citep{Jogee2005}. The rotation sense is set to clockwise, as the dust lanes along the bar are assumed to be leading. The dissipation rates were initially left at default values and only changed after the other parameters had been fixed. Table \ref{barmodel:preamble} lists the values of the potential parameters with which the model best reproduces our observations. The best fit was estimated by comparing the modelled channel maps (Fig. \ref{model_channelmap}) and intensity map (Fig. \ref{barmodel:figs} \textit{bottom, left}) against their observed counterparts by eye. In Appendix A we show the agreement of our best fit model with the CO observations at several different positions in the data cubes.
\subsection{Best fit model}
\begin{figure}
\resizebox{\hsize}{!}{\includegraphics{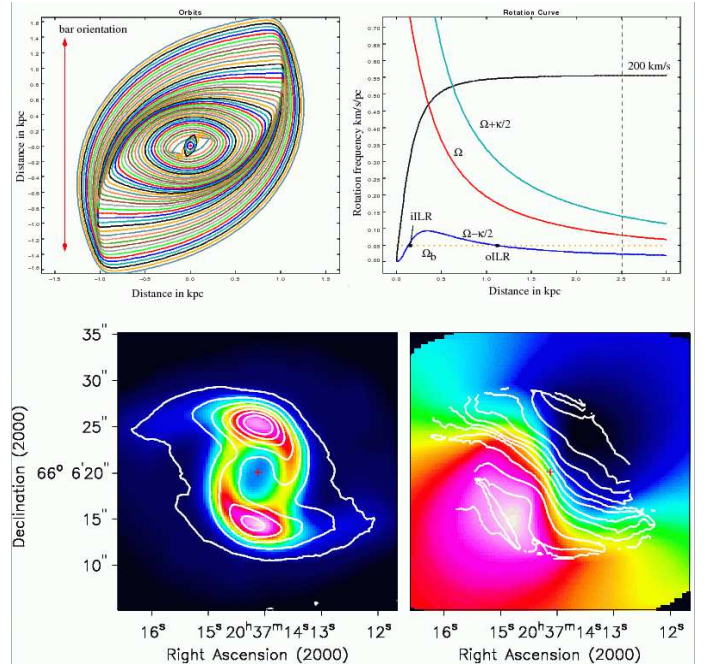}}
\caption{\textit{Top, left}: Deprojected orbit pattern in the inner 3.0\,kpc of the bar model. \textit{Top, right}: Resonance diagram. The rotation frequency $\Omega$ (red), $\Omega -\frac{\kappa}{2}$ (blue), $\Omega +\frac{\kappa}{2}$ (cyan) and the bar pattern speed (yellow dashed). The iILR at 160\,pc, oILR at 1150\,pc and CR at 4000\,pc (outside the displayed range) are indicated with black dots. Overlayed is the rotation curve, which approaches 200 km s$^{-1}$. The vertical dashed line indicates the outer edge of the models radial mass distribution. \textit{Bottom, left}: Integrated emission of the bar model with contours of CO(1-0) intensity map overlaid. The model emission has been computed at the CO(1-0) resolution. Contours run from 5$\sigma$ in steps of 20$\sigma$ (1$\sigma$ = 0.025 Jy beam$^{-1}$ km s$^{-1}$). The red cross indicates the position of the center. \textit{Bottom, right}: Velocity field of the bar potential model with contours of the CO(1-0) velocity field. Contours run from -200 to 200 km s$^{-1}$ in 25 km s$^{-1}$ steps.}
\label{barmodel:figs}
\end{figure}

Figure \ref{barmodel:figs} shows the orbits (\textit{top, left}) and rotation curve (\textit{top, right}) for our best model. The orbit pattern shows the change from $x_{1}$ to $x_{2}$ orbits, the main periodic orbits in a bar potential, starting at r$=$1.5\,kpc. The existence of two ILRs, denoted iILR (inner) and oILR (outer) in our model, can be seen from the double intersection of the $\Omega -\frac{\kappa}{2}$ curve with the pattern speed. Corotation is beyond 2.5\,kpc, which means the model does not diverge in the range we are studying here.

The presence of two ILRs in NGC\,6951 has also been found by both \citet{Rozas2002} and \citet{Perez2000}, who put them at 180\,pc and 900\,pc, and 180\,pc and 1100\,pc, respectively. Our CO observations cannot be fit by the model without 2 ILRs. Their locations are determined by the free parameters of the model; the pattern speed and the shape of the potential (defined by its characteristic length and velocity). From the resonance diagram in Fig. \ref{barmodel:figs} we find our ILRs to be at 160\,pc and 1150\,pc. These radii are close to the estimates reported by the earlier papers. The orientation of the bar perturbation (130$\degr$) in the model is also close to the value for the bar reported by \citet{Mulchaey1997}. This degree of consistency shows that the model is a good representation of the observations, and thus we conclude that the CO gas kinematics are indeed dominated by the large-scale stellar bar.

\begin{figure*}[htpb]
\centering
\includegraphics[width=17cm, angle=-90]{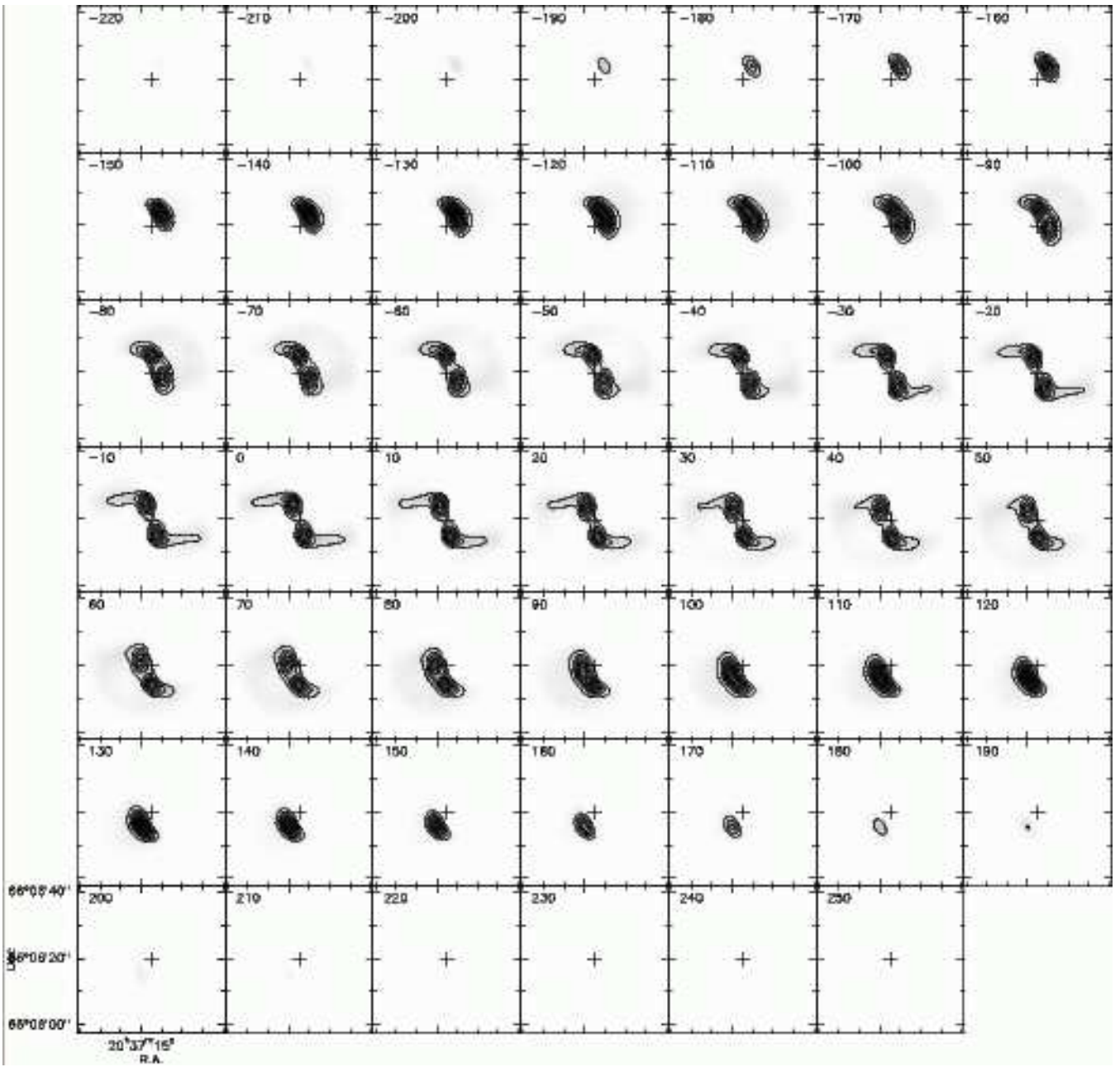}
\caption{Channel maps of our best model fit at the resolution of the CO(1-0) data. The FOV of each channel map is 44$\arcsec$ by 44$\arcsec$. The velocity indicated in the upper left corner is relative to the central channel (0 km s$^{-1}$).}
\label{model_channelmap}
\end{figure*}

The final model moment maps and channel maps are presented in Figs. \ref{barmodel:figs} (\textit{bottom}) and \ref{model_channelmap}. The model has been computed at the resolution of the CO(1-0) data. A comparison of Fig. \ref{model_channelmap} with the observations (Fig. \ref{channelmaps_co10}) shows good agreement. The model reproduces the straight gas lanes as well as the two peaks of emission in the ring in the individual channels from $-$30 km s$^{-1}$ to 40 km s$^{-1}$. The channels with emission in the model are only slightly fewer in number than in the observed channel maps. In the integrated emission map of Fig. \ref{barmodel:figs} (\textit{bottom, left}), the two spiral arms and the twin peaks are also well reproduced in this model. It is very encouraging that this bar potential model with its simple kinematics and intrinsic limitations fits the observations so well. Therefore, we are confident when we use the model velocity field to separate the two velocity components we detected in the data cubes.

In the channels, -30 km s$^{-1}$ to 40 km s$^{-1}$, the orientation of the two emission peaks is slightly more north-south in the model channel maps than in the observed one. From Fig. \ref{velomap21} we already saw that when we decompose the velocity field with two Gaussian fits, the velocity contours of the `bar'-component also become more north-south oriented.

In Sect. \ref{sec:morph} we reported the location of higher density gas traced by CO(2-1) and HCN further downstream in the spiral arms than  what is seen in the CO(1-0) line, which is mainly tracing lower-density gas. High emission peaks in both the CO(2-1) and HCN gas distribution occur away from the `twin peaks' that dominate the CO(1-0) distribution. We attemped to reproduce this pattern by also computing a best model at the resolution of the CO(2-1) data. The model gas distribution still agrees with the observations in its main characteristics; spiral arms and the twin peaks. However, the details, such as the southern intensity ridge and the second peak in the northern spiral, are no longer well reproduced. We point out that this model therefore reproduces a morphology closer to the CO(1-0) emission than either of the higher gas density maps mentioned above. A plausible explanation could be that the CO(2-1) and HCN distributions reflect the evolution of the molecular gas inside the circumnuclear ring itself, after the large-scale bar has brought the molecular gas there. As the molecular gas streams within the ring, cloud collapse can occur and/or continue, leading to higher molecular gas densities away from the contact points between ring and gas lanes.
\begin{table}[htpb]
\begin{minipage}{\columnwidth}
\caption{Bar potential model parameters}
\label{barmodel:preamble} 
\begin{tabular}{l c c}
\hline \hline
Parameter & Value & Initial Guess from data\\
\hline
Inclination (deg) & 46 & 46\\
Position Angle asc. node (deg) & 318 & 318\\
Char. Length (pc) \textbf{[r$_{p}$]} & 230 &\\
Char. Velocity (km s$^{-1}$) \textbf{[v$_{p}$]} & 200 &\\
Perturbation Azimuth (deg) & 130 & 84\\
Sense of Rotation & Clockwise & Clockwise\\
Bar Strength \textbf{[$\epsilon$]} & $-$0.035 &\\
Pattern Speed (km s$^{-1}$ kpc$^{-1}$) & 50 &\\
Rad. Diss. Rate (km s$^{-1}$ pc$^{-1}$) & 0.000 & 0.005\\
Az. Diss. Rate (km s$^{-1}$ pc$^{-1}$) & 0.020 & 0.002\\
\hline
\end{tabular}\\

\textbf{Notes:} The effect of errors in these parameters on our results is discussed at the end of Sect. \ref{sec:torque}.
\end{minipage}
\end{table}
\begin{table*}[htpb]
\caption{Bar potential model radial density parameters}
\label{barmodel:radii} 
\centering
\begin{tabular}{l c c c c c c c c c c}
\hline \hline
Radius (pc) & 20 & 450 & 500 & 600 & 700 & 800 & 900 & 1200 & 1500 & 2500\\
\hline
Col. Dens. (rel. scaling) & 0.25 & 0.2 & 0.6 & 0.7 & 0.6 & 0.6 & 0.3 & 0.1 & 0.1 & 0.02\\
Velocity Disp. (km s$^{-1}$) & 30 & 30 & 30 & 30 & 30 & 30 & 30 & 30 & 30 & 30\\
Scale Height (pc) & 10 & 10 & 10 & 10 & 10 & 10 & 10 & 10 & 10 & 10\\
\hline
\end{tabular}
\end{table*}
\section{Model derived results}\label{sec:torque}
The model described in Sect. \ref{sec:model} can also be used as the basis for a gravitational torque computation. The model's large-scale bar potential is known at each position, since the model is purely analytical. This makes it easy to derive the gravitational torque $\bar{T}(r,\phi)$ on the gas particles at each point, following the equation from \citet{Boone2007}:\\
\begin{equation}\label{eq:torque}
 \bar{T}(r,\phi) = -\epsilon\,v_{p}^{2}ln(1+\frac{r^{2}}{r_{p}^{2}})sin(2\phi)
\end{equation}
Here $\epsilon$ is the bar strength, $v_{p}$ the characteristic velocity and $r_{p}$ the characteristic length as set in the model. $\phi$ is the azimuthal angle with respect to the bar position angle.

Since the orbits for the gas particles are known, we define the net torque at a given radius as the time-average integrated torque over the orbit with this characteristic radius. We recall that the characteristic radius of an orbit is the radius this orbit would have without a bar (the orbit would then be circular). As mentioned in Sect. \ref{sec:model}, this approach is not completely self-consistent because when there is angular momentum change the orbits cannot be perfectly closed. Therefore, the gravitational torques computed here should be considered as a first-order approximation to the case in which the orbits are nearly closed. This is a valid approximation if the angular momentum loss is small. Using the simple argument that $\frac{dL}{L}$ is of order $\frac{T}{v^{2}}$, where $v$ is the characteristic velocity of the potential v$=$200 km s$^{-1}$, we can see that the relative amplitude of the angular momentum loss per orbit is of order $\sim 0.0125-0.05$. Therefore, it takes at least 20 orbital times for the gas to fully lose its angular momentum.

\begin{figure*}[htpb]
\centering
\includegraphics[width=17cm]{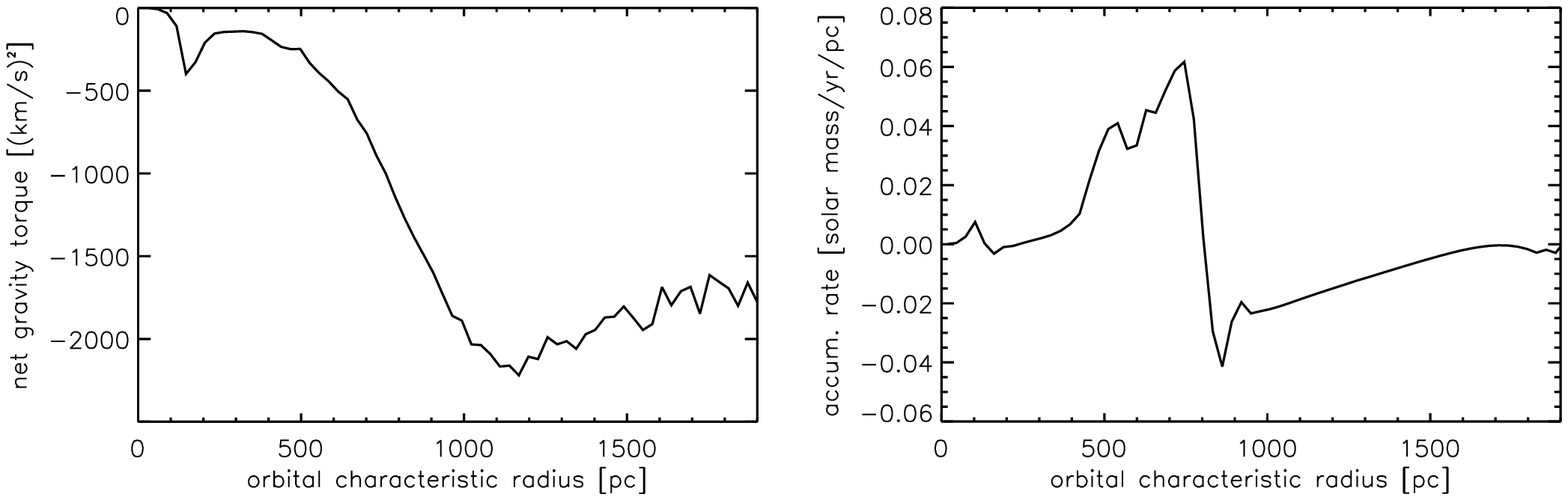}
\caption{\textit{Left}: Net gravity torque derived from the best fit model. \textit{Right}: Gas accumulation rate as a function of radius. The orbital characteristic radius corresponds to the radius of circular orbits that a given gas particle would maintain if unperturbed by the model's large-scale bar.}
\label{torques_outputfigs}
\end{figure*}

Figure \ref{torques_outputfigs} (\textit{left}) shows the net gravity torque for each orbit in the model. On the horizontal axis we plot the characteristic radius of each orbit. The net gravity torque is negative over the entire radial range studied here. From 1.9\,kpc down to 1.0\,kpc the torque stays somewhat constant at values of $-$2000(km/s)$^{2}$. There is a slight upturn at larger radii. This is the region of the straight gas lanes. The large value indicates that the bar is easily able to transport gas inward to the gas spiral arms. The torques then increase at an almost constant rate down to a radius of 400\,pc. We can understand this from Fig. \ref{barmodel:figs} (\textit{top, left}): the orbit position angle with respect to the bar stays almost constant (at its maximum value) and the orbit elongation decreases with decreasing radius. A second, smaller, minimum is present at r$=$150\,pc, at the iILR. The net gravity torque becomes zero within the 50\,pc radius. A close inspection at the model orbits in the inner 300\,pc shows that the orbit orientation changes again from $x_{2}$ to $x_{1}$.

The mass accumulation rate that we can calculate from the gravitational torques also depends, in contrast to the gravitational torque per se, no longer solely on the analytical potential, but also on the radial gas mass distribution in the model. Fig. \ref{torques_outputfigs} (\textit{right}) shows the mass accumulation rate for this model. The variation in the mass accumulation rate indicates that gas must accumulate on some orbits and be depleted from others. That the gravitational torques drive the gas inward is also evident from this curve. At radii larger than 800\,pc gas is being depleted and the accumulation rate is negative. These are the radii where the gas is located in the straight gas lanes. From 400\,pc to 800\,pc radius, the accumulation rate becomes positive. These are the radii at which we find the gas spiral arms and the gas ring, which are separated into two peaks in the figure. In the radius range 400-1000\,pc we already saw from the torque plot that the torques decrease with decreasing radius. As the radius decreases the torques are less strong, so less gas is being moved inwards at each radial bin and the gas can accumulate. Inside 400\,pc we see a sharp positive peak, at 150\,pc. This is the radius at which we found a local minimum in the gravitional torques. The feature is confined to only a very limited radial range, but indicates that some accumulation may still happening inside the ring.

Mass accumulation within 1.5\,kpc in radius, i.e. integrated over this radial range, is occuring at a rate of $\sim$+2.0 M$_{\sun}$yr$^{-1}$ and mostly in the two spiral arms. From our observations we know that the total amount of molecular gas within this radius is about 1.6$\times$10$^{9}M_{\sun}$. Therefore, at our derived accumulation rate, all matter in the spiral arms must have been brought there within the last 0.8 Gyr. However this is a lower limit to the age of the ring, since it assumes that no gas has been turned into stars and that there is always enough gas available at larger radii to be driven inward.

Errors in the gravitational torque computation can come from six model parameters; the characteristic length and velocity of the potential, the bar pattern speed, the bar strength and the dissipation factors. The first three influence the locations of the resonances as well as the values of the net gravity torques. A 10\% difference in either of these parameters can cause up to a $\sim $20\% difference in the locations of the gravitational torque minima. The change in value is 5\% for the characteristic length and 20\% for the characteristic velocity and pattern speed. The net gravity torque scales linearly with bar strength. The dissipation factors only influence the strength of the gravitational torques. A 10\% difference in these factors leads to a decrease of $\approx$ 25\% in the net gravity torques. However, such changes to the model parameters lead to significantly worse models. Therefore, the uncertainties in the results we derive here for the net gravity torque and the mass accumulation rate are within the relative errors given here.
\section{Discussion}\label{sec:discuss}
\subsection{Gravitational torques}\label{sec:torquecomp}
We can now compare different gravitational torque measurements we obtain using either the `torque map' method introduced by \citet[][; hereafter: GB05]{Garcia2005} \citep[see also][; hereafter: H09]{Haan2009}, and the analytical computation possible for the large-scale bar model following \citet{Boone2007}.

\begin{figure}[htpb]
\resizebox{\hsize}{!}{\includegraphics{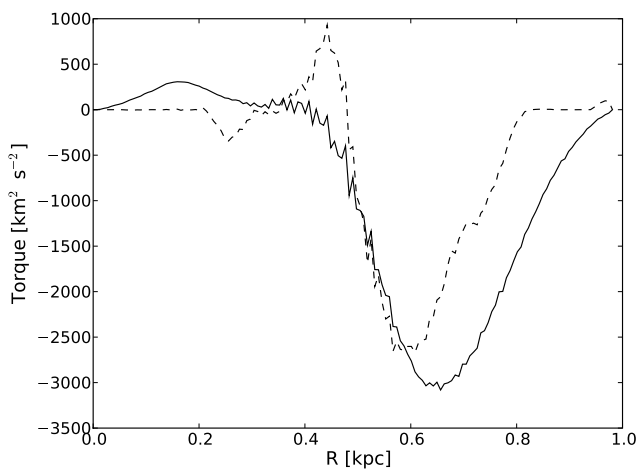}}
\caption{Gravitational torque curve derived from the PdBI-only data ((\textit{dashed line}) reproduced from \citet{Haan2009}). Gravitational torque curve derived from the PdBI+30m data (\textit{solid line}) using the same method. The addition of the 30m data has a large influence on the torque values, especially inside 500\,pc.}
\label{withwithout}
\end{figure}

First, we investigate how the inclusion of the 30m data changes the results obtained with the `torque map' method. Using the code PyPot of H09 we have derived the gravitational torques for our PdBI+30m data cube. PyPot uses NIR high-resolution images to evaluate the stellar potential. The gravitational torque curves for PdBI-only (\textit{dashed line}) and PdBI+30m (\textit{solid line}) data are shown in Fig. \ref{withwithout}. As can be seen from the figure, the inclusion of the 30m data leads to a change in the gravitational torque budget, especially in the region inside 500\,pc. The high positive torques derived at 400\,pc in the PdBI-only maps disappear in the PdBI+30m result. This can be explained by the larger amount of diffuse molecular gas that is recovered in the PdBI+30m data, which lowers the high torque barrier at $\sim$200\,pc found by GB05 (see their Fig. 12) and H09. The spiral arms become more prominent with the inclusion of the 30m data, so they contribute more to the torque budget at larger radii, shifting the minimum to larger radii. This test shows that sampling the line emission on all spatial scales is important when computing gravitational torques from observations as the absolute values can vary significantly while the overall shape is preserved.

Now, we turn to the comparison of the `torque map' result with the analytical computation. We are interested in the torques as we want to compute the angular momentum change in the gas (whether the gas is in- or outflowing). In the analytical case, we can simply integrate the net torque over gas orbits from the bar model potential (which is a simplification of the true underlying potential). This integration gives the angular momentum loss of a gas particle in one orbit. In the case of `torque maps' it is often argued that it is impossible to know the exact orbits from observations. So, a statistical argument is used in the `torque map' method. The assumption is that the observed gas distribution can be linked to the time spent by the gas along the orbits. The resulting torque is the gas column density weighted average of the local torques averaged per radial bin.

There seem to be significant differences between the results form both methods. In the `torque map' result we find the strongest (negative) torques at a radius of $\sim$650\,pc. For the analytical computation the most negative torques are approximately at a radius of r$\sim$1200\,pc. As we saw from the intensity maps (Fig. \ref{CO10intens}), we observe very little gas in the straight gas lanes (r$\gtrsim$1.0\,kpc). We know, however, that the large-scale stellar bar has a radius of 3.0\,kpc \citep{Mulchaey1997}. At 1\,kpc the large-scale bar still exerts significant torques on the gas, but the `torque map' method being very sensitive to the actual observed gas distribution would not show that since there is so little gas, hence the earlier upturn.  There is also a difference in the absolute values of the strongest negative torques between the `torque map' method and the analytical computation. This inequality is due to the difference between computing the net torque (model) and averaging the torques (maps).

In the inner 300\,pc, another difference is seen between the `torque map' method and the analytic computation. In our analytical computation we see a second small dip in the torques, while in the `torque map` result we find positive torques. By construction, it is impossible in our model for the torques to change sign. Further, GB05 have found that there is a secondary gravitational component, a nuclear stellar oval, that plays a role in the torque budget here. No such secondary component has been included in our model.

Despite these differences, the comparison between our mass accumulation rate in Fig. \ref{torques_outputfigs} and the mass accumulation rate found by H09 (see their Fig. 12) shows little difference. We find values not more than a factor 2 larger than theirs. This agreement shows that our large-scale bar potential model and the subsequent analytical computation give a good representation of the true gas transportation in this region, at least from the outer disk down to the circumnuclear ring.
\subsection{CO gas flow inside the ring}\label{sec:nucflow}
Another aim of this paper is to see if a large-scale bar can explain the gas distribution and kinematics over the entire 1.5\,kpc radial range. For all radii down to the ring the model represents our observations very well. Inside the ring we found negative torques in our model, but positive torques based on H09's `torque map' method. Does the influence of the large-scale stellar bar extend further in? As a first test we looked for evidence of gas spiral arms inside the ring, as predicted by the model of \citet{Englmaier2000} and \citet{Maciejewski2004II}. In Fig. \ref{JH-overlay} we see some indication of spiral structure in the dust. In the observed CO morphology we do not, nor do we find such evidence when we decompose the velocity field \citep{Fathi2005,Glen2010}. The latter method is sensitive to even a small arm/inter-arm density contrast. Based on this outcome, it seems that the CO gas is mainly on stable orbits in this region, which implies that it is not inflowing. However, we did find other evidence of CO gas that is potentially not on stable $x_{2}$ orbits; the gas-bridge and the double CO emission peaks in the line profiles (Figs. \ref{gasbridge} and \ref{velomap21}, Sect. \ref{sec:kin}).

The primary component of the double peak emission inside the ring (Fig. \ref{velomap21}, middle right) shows velocities indicative of stable orbits. The line-of-sight velocity of the `second peak' CO component (located within the white ellipse in the bottom right panel of Fig. \ref{velomap21}) is 70-90 km s$^{-1}$, which corresponds to a velocity difference from the primary CO component at these coordinates of $\Delta$v $\sim$ 80 km s$^{-1}$. It is unlikely that we have a simple superposition of two gas clouds, both on stable orbits. After separating the flux in this area into the two velocity components, we derive an H$_{2}$ mass for this `second peak' CO component of 1.8$\times$10$^{6}$M$_{\sun}$.

To understand if this CO component might still be related to gas spiral arms induced by the large-scale bar, the gas flow direction of the `second peak' component and whether it is located in the disk of the galaxy must be ascertained. The first issue is straightforward. As \citet{Storchi2007} have concluded the near side of the galaxy is to the southwest and the far side to the northeast. Thus, the positive velocity of this CO component, which is south-south-west of the nucleus, indicates streaming toward the nucleus if the component is in or in front of the disk, and outflow if it is behind the disk. Outflows are usually biconical. If this component is part of an outflow, seen behind the disk, we would expect to see a second blue-shifted component from the outflow in front of the disk, which is not the case. Nor do we really expect an outflow, since this galaxy does not have a strong AGN or starburst near its nucleus. It is thus unlikely that this CO component is located behind the disk and outflowing.

If the component is inflowing, it is very unlikely that it has any significant scale height above the disk. NGC\,6951 is an isolated galaxy that has been dynamically undisturbed for at least 1 Gyr. This history means that the origin of the CO component has to be internal. It could be that some intense star formation in the ring expelled CO from the ring. We do see a gap in the CO ring close to the component. However, the low velocity dispersion of the CO gas in the component makes it unlikely that the CO component was heated due to star formation, became unbound and was pushed high above the disk and is now falling back. Nor do we see evidence of a recent starforming event at that location.

Thus, it seems most likely that the CO component is inflowing and located in the disk, bringing us back to the question of whether the component's motion can be explained by large-scale bar induced gas spiral arms, even though our initial tests were negative. \citet{Storchi2007} claim to see evidence of two such spiral arms in H$\alpha$ line emission. The location of their H$\alpha$ spiral arms does not match the location of our CO component however. Their derived velocity of $\pm$ 20 km s$^{-1}$ is also at least a factor 3 smaller than that of our CO component. If there are really gas spiral arms in NGC\,6951, this discrepancy is very unexpected. CO should be a better tracer of such gas spiral arms, since it is the dominant gas tracer in this region. So it seems that the large-scale bar does not induce further gas inflow in this galaxy and that this CO component is inflowing due to some other reason.

The CO component might be connected to the stellar oval found by \citet{Garcia2005} inside the circumnuclear ring. The position angle of this oval is $\sim$ 66\degr. Unlike the H$\alpha$ spiral arms, its orientation matches well with the location of the CO component. If we draw a line through the dynamical center with the oval's PA, the CO component lies almost completely at the southern leading side of the oval (Fig. \ref{velomap21}, \textit{bottom left}). A stellar oval is in essence a `fat' bar and can drive gas inward similar to the large-scale stellar bar on larger scales (i.e. the bars-within-bars scenario). The location of the CO component with respect to the oval is correct for this scenario, although we cannot conclude inflow from our gravitational torque results. The reason why the influence of the stellar oval is not more prominent in the gas might be that its bar strength and thus the torque it exerts on the gas is low \citep[][; their Fig. 8b]{Garcia2005} and the amount of gas reaching the influence radius of the oval is limited by the circumnuclear ring and its associated star formation. The CO gas we detected in the gas bridge lies in the region where it would feel positive torques due to the nuclear oval; in the averaged torque budget, as shown in Fig. \ref{withwithout} this gas can dominate and mask the smaller amount of inflowing gas at these radii. Alternatively, viscosity torques may play a more significant role here, as was also dicussed by \citet{Garcia2005}.

This CO component could be one of the last links in the chain of gas inflow down to the center. The large-scale bar drives most gas down to the ring. Based on the gravitational torques, this is not the smallest possible radius-- only the mass accumulation rate becomes small. So, it is possible for gas to move further in, reach the sphere of influence of the stellar oval or experience viscosity torques, which brings it further toward the nucleus. The CO component we observe does not completely reach the dynamical center. However, spatially and kinematically, it connects to the central HCN detection of \citet{Krips2007}. The HCN extends over $\sim$0.5$\arcsec$ around the dynamical center, and our CO component extends to about this radius from the center. The detected HCN also has a detected velocity range of $\pm$ 70 km s$^{-1}$, with the positive side where the CO component is, which also has a velocity of 70 km s$^{-1}$. This concentration potentially completes the chain of gas inflow.
\section{Summary and conclusions}
We have presented high-resolution $^{12}$CO(1-0) and $^{12}$CO(2-1) maps of the central region of NGC\,6951 using the IRAM PdBI and 30m telescopes. This galaxy exhibits significant indirect evidence for recent gas inflow. It has a large-scale bar, a circumnuclear SB ring, and an AGN. Molecular gas, as traced by CO, is the dominant phase of the ISM in these regions.

Investigation of the CO data cubes shows that the gas distribution can be well explained by gas response to a large-scale stellar bar. We further quantify this scenario by reproducing the observations using a model where the gas is solely responding to a large-scale stellar bar. The model used here has the advantage that it provides direct predictions for the radial motions of the gas. Our best model is able to reproduce the main characteristics of the observed CO morphology and kinematics.

Gravity torques and gas mass accumulation rates that follow from this model were computed and compared to previous gravity torque maps by \citet{Garcia2005} and \citet{Haan2009}. Their method is complementary to ours and we explain the differences between the two results. As expected, we find that the large-scale bar effectively transports gas inward to r$\sim$400\,pc, but it no longer dominates the gravitational torque budget on smaller radii.

The stellar oval reported by \citet{Garcia2005} is a likely candidate to take over gas transport inside the circumnuclear ring. Detailed investigation of the data cubes revealed double-peaked CO profiles at several positions inside the gas ring. Most notably, we detect a second CO complex near the dynamical center, with a velocity offset of $\Delta$v $\sim$ 80 km s$^{-1}$, which is not on a stable orbit. From simple geometric and physical arguments, we conclude that this cloud is in the disk of the galaxy, is inflowing toward the nucleus, is likely driven by the stellar oval, and is indeed the last step in bringing gas close to the nucleus.

\begin{acknowledgements}
T. vd L. acknowledges DFG-funding (grant SCHI 536/2-3).
\end{acknowledgements}

\bibliographystyle{aa}
\bibliography{/home/vdlaan/Papers/Paperbib}

\onecolumn
\appendix
\section{Models' goodness of fit}
\begin{figure*}[htpb]
\includegraphics[width=17cm]{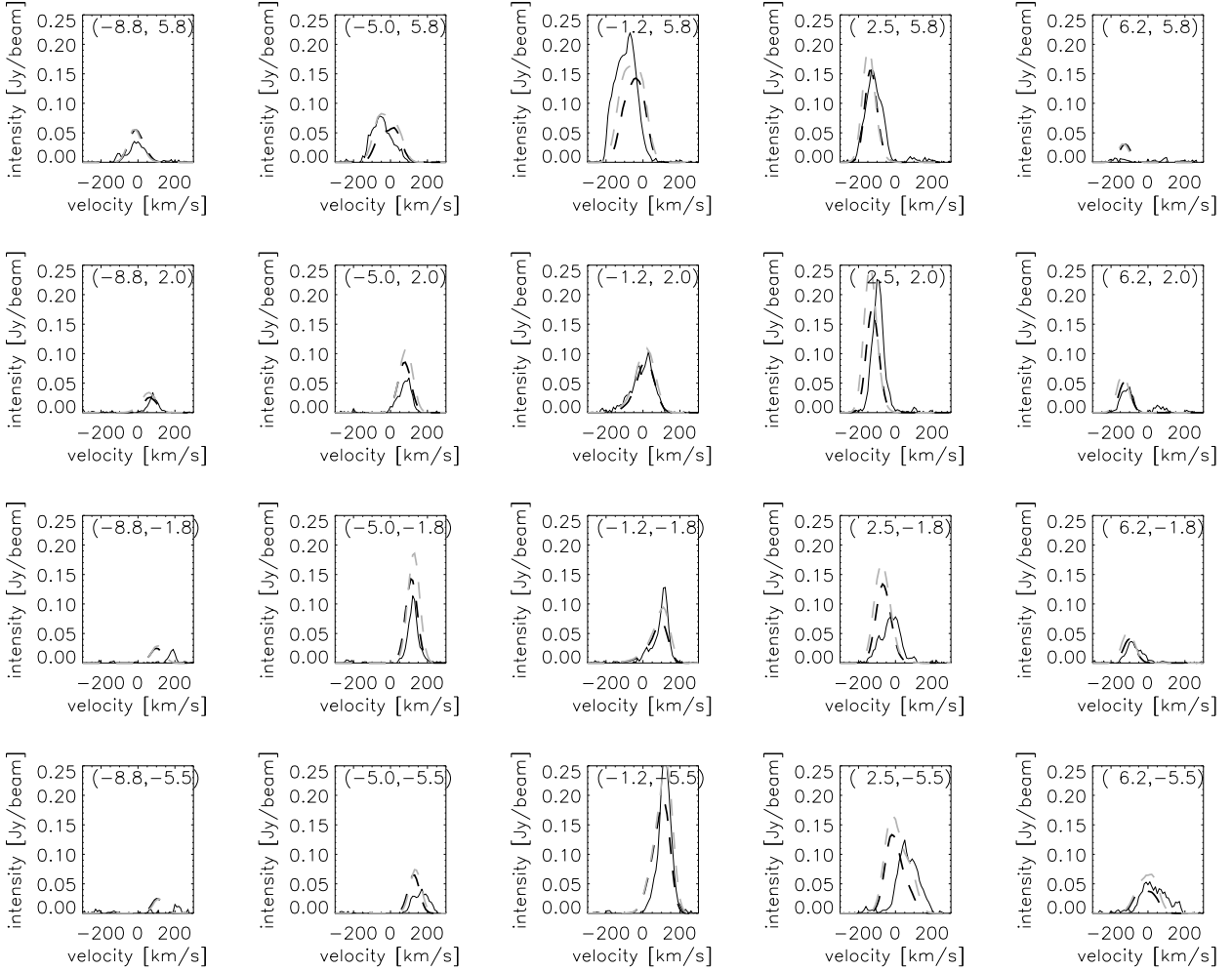}
\caption{CO(1-0) emission spectra at different positions in the observed data cube (solid line), the best model cube (black dashed line) and a rejected model cube (grey dashed line). The spatial positions in the galaxy for each spectrum, in arcseconds, are given in the subpanels. The model shows a good agreement with the observations. The positions (-1.2,5.8) and (-1.2,-5.5) correspond to the `twin peaks' described in Sect. \ref{sec:morph}}
\label{fig:goodfityeah}
\end{figure*}

In Sect. \ref{sec:model} we constructed a bar model based on an analytic potential, and derived a corresponding model data cube. The best model was estimated by comparing the model's channel maps and intensity map against their observed counterparts. In Fig. \ref{fig:goodfityeah} we present overlays of the observed and modelled spectra at different positions in the cubes to further show the goodness of fit.

Fig. \ref{fig:goodfityeah} shows that the model (black dashed lines) follows the observed line emission (solid lines) very well. There are some local differences in the intensity. These differences occur mainly at the locations of the `twin peaks', (-1.2,5.8) and (-1.2,-5.5). There the observed emission is larger than we reproduce with the model. A rejected model, with slightly different values for the parameters char. length (210\,pc), char. velocity (220 km s$^{-1}$), bar strength (-0.030) and bar pattern speed (60 km s$^{-1}$ kpc$^{-1}$), is shown with grey dashed lines. This model was rejected because it overpredicts the intensity in the ring.

\end{document}